# WebCrack: Dynamic Dictionary Adjustment for Web Weak Password Detection based on Blasting Response Event Discrimination

Xiang Long
*School of Cyber Science and Engineering*
*Huazhong University of Science and Technology*
Wuhan, China
longxiang@hust.edu.cn

Yan Huang
*School of Artificial Intelligence and Automation*
*Huazhong University of Science and Technology*
Wuhan, China
platanus@hust.edu.cn

Zhendong Liu
*School of Cyber Science and Engineering*
*Huazhong University of Science and Technology*
Wuhan, China
949833447@qq.com

Lansheng Han*
*School of Cyber Science and Engineering*
*Huazhong University of Science and Technology*
Wuhan, China
hanlansheng@hust.edu.cn

Haili Sun
*School of Cyber Science and Engineering*
*Huazhong University of Science and Technology*
Wuhan, China
hailisun@hust.edu.cn

Jingyuan He
*School of Cyber Science and Engineering*
*Huazhong University of Science and Technology*
Wuhan, China
3452108897@qq.com

*Abstract*—The feature diversity of different web systems in page elements, submission contents and return information makes it difficult to detect weak password automatically. To solve this problem, multi-factor correlation detection method—as integrated in the DBKER algorithm—is proposed to achieve automatic detection of web weak passwords and universal passwords. It generates password dictionaries based on PCFG algorithm, proposes to judge blasting result via 4 steps with traditional static keyword features and dynamic page feature information. Then the blasting failure events are discriminated and the usernames are blasted based on response time. Thereafter the weak password dictionary is dynamically adjusted according to the hints provided by the response failure page. Based on the algorithm, this paper implements a detection system named WebCrack. Experimental results of two blasting tests on DedeCMS and Discuz! systems as well as a random backend test show that the proposed method can detect weak passwords and universal passwords of various web systems with an average accuracy rate of about 93.75%, providing security advisories for users' password settings with strong practicability.

*Keywords—web security, weak password detection, universal password, automatic login, dynamic dictionary adjustment*

I. INTRODUCTION

Nowadays, the web has brought us great convenience and facilitate the dissemination of information on the Internet. However, it also brings a series of security problems. The large amount of individual and enterprise private data from web backend has become the focus of hacker attacks [1][2], with weak password attack [3][4] being one of the most direct and effective attacks.

Dinei et al. studied the password habits of 500,000 users for three months [5] in 2007 and found that nearly 70% of users were using weak passwords. In 2011, Cui et al. used Nmap to conduct a weak password detection study on HTTP and Telnet service hosts around the world [6] and found 1.1 million devices with weak passwords. In 2015, Patton et al. [7] detected weak password on multiple types of the Internet of Things (IoT) devices with the Shodan search engine and found that the rate of weak password vulnerability reached to more than 40% on some devices. Given their vulnerability, weak passwords have brought huge security risks to the Internet, and therefore, how to detect weak passwords in web systems has become particularly important.

This also raises the question though, is a web management system with complex passwords secure enough? Actually, that is not the case. In addition to weak passwords, the universal password vulnerability is also an aspect that security workers need to pay attention to. The reason for this vulnerability is that the SQL keywords of the accepted parameters are not filtered in the login backend, leading to SQL injection [8]. This means that an attacker can use injection statements such as "or" a "=" a, a 'or' 1 = 1-- to enter the management backend without any password, making the password system virtually useless.

Due to the complexity of the web system [9], it is hard to judge the login status with the available static feature values that are specific and limited. Therefore, most security workers can only perform cumbersome manual detection for weak passwords [10]. The disadvantages of manual analysis are not obvious for those cases with a small number of test objects. However, as the number of tests grows, administrators often need to detect weak passwords in batches. At this point, manual analysis will lead to extremely low efficiency and time-consuming, thus it is highly necessary to design a general weak password detection algorithm as an alternative.

The main contributions of this paper are listed as follows:

(1) Propose a new automatic detection method for weak password and universal password vulnerability of the web system, called DBKER. It judges blasting result via 4 steps. Then the blasting failure events are discriminated and the usernames are blasted based on response time. Thereafter the weak password dictionary is dynamically adjusted according to the hints provided by the response failure page.

(2) Based on this algorithm, a detection system model – WebCrack is implemented. In addition, modules such as random headers, dynamic password generation and custom blasting rule expansion are added to reduce the probability of the detection algorithm being blocked by the firewall and to improve the recognition rate.

(3) Two blasting tests on DedeCMS and Discuz! systems as well as a weak password detection experiment on 14,185 randomly selected backend addresses were carried out, the results were analyzed and compared.

## II. RELATED WORK

In view of the weak password detection problem, there have been some related research works. Tian et al. proposed a method for weak web password detection based on static analysis of web pages in 2016 [11], which supports batch web weak password detection. In 2017, for the automatic detection of XSS vulnerabilities, Chen et al. proposed a method using simulated login to detect XSS for those cases where further testing could only be performed after login [12]. In 2018, Xu et al. designed a web weak password detection system AWKD [13] for Internet of Things, which also supports automatic capture and detection of IoT devices. TideSec security team released the open source project "web_pwd_common_crack[1]" on github in 2019, which contains an universal web weak password cracking script that can detect whether there is a weak password vulnerability in the backend without a verification code.

The characteristics of passwords and PINs chosen by Chinese and non-Chinese users have been studied by [14][15], The composition pattern of personal password is studied in depth. Different from such methods, we use the open password dictionary data set, do training on the open data set based on PCFG algorithm and then automatically generate weak password dictionary, and can also learn different password composition patterns. Ming et al.[16] regard a password as a composition of several chunks, where a chunk is a sequence of related characters that appear together frequently, to model passwords. Patrick et al.[17] develop an efficient distributed method for calculating how effectively several heuristic password-guessing algorithms guess passwords. Teng et al.[18] conduct password guessing based on recurrent neural networks and generative adversarial networks. Some other works[19][20][21] improve the blasting prevention ability of system login by studying and setting different password strength.

---

[1] TideSec/web_pwd_common_crack: Common Web Weak Password Cracking Script: https://github.com/TideSec/web_pwd_common_crack.

While the previous work has made some important contributions to the automatic detection of weak passwords, there remain some obvious shortcomings as well. For instance, the detection algorithm proposed by Tian et al. lacks detailed description of how to judge the correctness, nor does it check the validation of the final result [11]. Although the web system of Internet of Things devices has some similarities with the common web systems, the latter have more styles and are more complex. Thus the system designed by Xu et al. [13] works on IoT devices, but is not necessarily applicable to common web systems due to some complications. While the "web_pwd_common_crack" tool is simple to use and is open source, experimental results show that it has high false positives and low detection accuracy. In addition, none of the above systems or models can detect universal password vulnerabilities or support user-defined blasting rules, as well as lacking scalability.

To solve the above problems, this paper proposes a set of detection method combining traditional static and dynamic page features, i.e. DBKER (Dynamic, Blacklist, Login Key, Error Length, Recheck) algorithm. Different from existing works, it generates password dictionaries based on PCFG altgorithm, proposes to judge blasting result via 4 steps with traditional static keyword features and dynamic page feature information. Then the blasting failure events are discriminated and the usernames are blasted based on response time. Thereafter the weak password dictionary is dynamically adjusted according to the hints provided by the response failure page. Based on this algorithm, a system called WebCrack is designed and implemented, which can be used to detect backend weak password and universal password.

III. DBKER Blasting Algorithm

In this section, we analyze and classify the web blasting events at first, then propose an algorithm named DBKER to conduct blasting, judge the blasting failure and dynamically adjust the dictionary based on the response page discrimination. Finally, we calculate the detection capability of the algorithm for each event and the false positive rate overall. The workflow of the proposed web page blasting algorithm DBKER is shown in Fig. 1.

A. Blasting event analysis

When conducting website blasting, the input passwords can be divided into three different types: wrong passwords, correct passwords and universal passwords. Among them, the passwords that can log in normally belongs to correct passwords, the ones that do not cause firewall blocking but cannot log in are regarded as wrong passwords, otherwise belong to universal passwords.

According to the final position after the redirected with code 302[2], the response pages are divided into original pages and new pages. From the content of the pages, the response pages are divided into password error prompt pages, password correct prompt pages, prompt pages exceeding the maximum number of password errors, firewall blocking pages, and background pages. To facilitate the discussion below, according to experimental statistics, the blasting response events can be categorized into the following 9 situations, which are recorded as Events 1–9 respectively. The pages that do not meet this general rule, such as incorrectly positioned login boxes or intermittently unstable pages, etc., are considered as "Other" cases. Among them, Events 1–7 are interference events, and Events 8 and 9 are login success events that need to be filtered out.

■ Event 1: Enter wrong password, original page, no prompt.

■ Event 2: Enter wrong password, original page, prompt the password error page.

■ Event 3: Enter wrong password, original page, prompt for exceeding the maximum number of password errors.

■ Event 4: Enter wrong password, new page,

---

[2] The HTTP response status code 302 is not an error code and actually means URL redirection, i.e., the URL accessed by the current connection is redirected to the new URL.

prompt password error page.

- Event 5: Enter wrong password, new page, prompt for exceeding the maximum number of password errors.
- Event 6: Enter universal password, new page, prompt the firewall blocking page.
- Event 7: Other.
- Event 8: Enter correct password, new page, prompt the password correct page.
- Event 9: Enter correct password, new page, enter the background.

*B. Determine page stability and obtain error length*

Before web page blasting, we need to determine page stability and obtain error length at first (refer to the preprocessing detail in the supplemental material 1). Here we define concept "page stability" as: return the same result for the same packet. If the page is stable, it indicates the page returns consistent results for wrong passwords, which can be regarded as the characteristics of our judgment. If the page is inconsistent, it means that the page is unstable, and results will be dynamically returned. For this case, the system cannot judge and exits the blasting test.

In detail, the system will first send two incorrect passwords to determine the stability of the page. If the lengths of the two returned pages are equal, the page is considered stable. Meanwhile, the total length of the page returned at this time, i.e. the Error Length (hereinafter referred to as EL), is recorded as the benchmark for the judgment in the third step. If unstable, the program exits.

In addition, some management systems will allocate a cookie [23] to record the login times of users at the first request, so the system will first request the backend page, which we call a pre-request, to save this cookie value to avoid the interference of the judgement due to the length of set-cookie in the response header. This was tested in experiment 1 and 2.

*C. Dictionary generation based on PCFG*

In practical applications, in addition to setting their own personal information, users may also refer to the visible information in the website or the information known by multiple account managers. Attackers usually cannot directly access users' personal information, but directly public information on websites can be easily obtained. It is better to find and collect the public information related to the user's login network environment, and then conduct the user password guess attack.

The existing directed password guessing method only considers some personally identifiable information in the leaked password set, such as name, birthday, email address and so on. In the actual work environment, if the account is not completely private property, users tend to add public information when setting passwords, such as the public telephone number, the name or abbreviation of the unit or organization they belong to, the name of the area where the unit belongs to and other well-known information. From the attacker's point of view, this information is available on several pages of the site. If the attacker has access to such public information, it can significantly improve the success rate of password guessing. This customized information is not available from weak password dictionaries or previously disclosed password data sets, but can be easily obtained through manual look-up or web crawler methods.

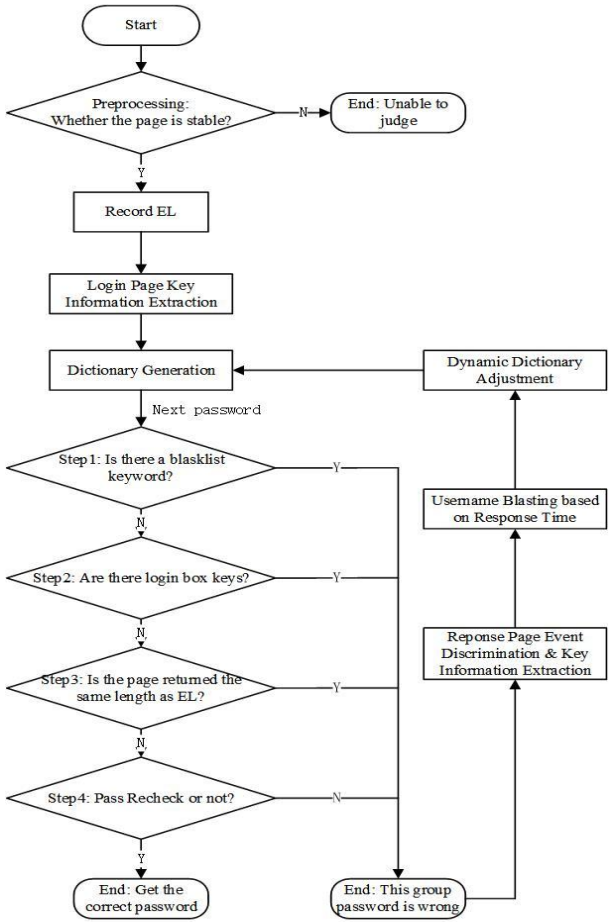

Fig. 1. The overall workflow of the DBKER algorithm

Therefore, after preprocessing, we extract key information from the origin login page at first, and then generate dictionaries for username and password guessing based on PCFG algorithm. The domain names, IP addresses and company names may reveal some clues about users and organizations, and we use the information to generate password dictionaries.

PCFG algorithm is a walking password guessing attack algorithm proposed by Weir et al. based on random context-free grammar. It sets a probability value for each structural rule and field in the corpus, and then obtains the probability of generating a password. PCFG algorithm regards the input password as several independent segments, which include three types: alphabetic segment, numeric segment and special character segment, represented by $L$, $D$ and $S$ respectively. Each field has a numeric subscript to indicate the length of the segment. The password can be divided according to the preceding rules. For example, the password "password6789!". The value can be divided into three characters: alphabetic segment $L_8$ : "password", numeric segment $D_4$ :"6789", and special character segment $S_1$ :"!", the structure of the whole password can be expressed as $L_8D_4S_1$. PCFG algorithm includes training stage and guess generation stage.

The input of the training stage is the password set to be analyzed, generally the existing public password set. The output of this stage consists of two parts, namely, the frequency table $\Sigma_1$ of password structure and the frequency table $\Sigma_2$ of password characters. For example, the password structure $L_8D_4S_1$ of "password6789!" will appear in the password structure frequency table $\Sigma_1$, and the frequency of the structure can be calculated together with other passwords of the same structure. The password frequency table $\Sigma_2$ calculates the frequency of these partitions, including the frequency of "password" in the 8-character L segment, "6789" in the 4-character D segment, and the frequency of "!" in the S segment.

The input of the guessing generation stage is the password set whose probability needs to be calculated, and the output is the probability value of each password in the password set. The principle of guessing is to generate a guessing set with probability by multiplying probabilities according to the password structure frequency table and password character frequency table obtained in the previous stage. For example, guess the password "password6789! The calculation process of can be expressed as the following formula:

$$P(\text{password6789!}) = $$
$$P(S \rightarrow L_8D_4S_1) \times P(L_8 \rightarrow password)$$
$$\times P(D_4 \rightarrow 6789) \times P(S_1 \rightarrow !)$$

According to the above formula, we can obtain the generation probability of password "password6789!". By ordering the generated password structure frequency table $\Sigma_1$ and password character frequency table $\Sigma_2$ in descending order, the guessing set used in password guessing attack can be obtained.

D. *Blasting result judgement*

  *1) Step 1: Keyword Blacklist Detection*

Blacklist detection is a static feature detection method. By collecting and organizing common login failure keywords, most error password pages can be filtered. In general, there are mainly three types of keywords:

(1) Keywords prompting the wrong password.

(2) Keywords prompting that exceed the maximum number of attempts.

(3) Keywords prompting blocked by the firewall.

Passing the blacklist keyword detection alone does not guarantee that the password will always be correct. The reason for this is that (1) we cannot collect all the wrong keywords, and (2) most of the keywords are only English and Chinese keywords, so for websites in other languages, this step does not work.

When there are keywords in the blacklist on the returned page, it means that the password is incorrect. Then, the next set of password attempts is made. If the password tried does not exist, proceed to the next step. This step can filter interference Events 2, 3, 4, 5, 6. The filtering capability depends on whether the blacklist keywords are complete.

*2) Step 2: Determine whether the page after jump contains the login box key*

In the case of Events 1, 2, 3, if the password entered is incorrect, the login box of the web system will continue to exist to allow the user to re-enter the password; if the password is correct, it will enter the management backend and the login box disappears. So the login box can be used as the criterion for the judgment. When identifying the parameters, all the parameters in the login box will be extracted, including the key name of the account and the password that needs to be blasted. At this point, their key names are marked and recorded as user_key and pass_key (hereinafter collectively referred to as keys), then compared with the response page. If the keys still exist, this group of passwords is incorrect, proceed to the next judgment. If not, proceed to the next step.

Here a question arises: whether to compare with the page at the 302 jump or the page after the 302 jump? There are many ways to jump, and many kinds of prompt information, but in the end, it still has to fall on the final information returned. This means that the content of the page at the 302 jump cannot be used as a basis for judgment. Therefore, whether the recorded keys exist in the page after the 302 jump needs to be taken into consideration. Unless otherwise specified, this article focuses on the content of the page after the 302 jump by default.

*3) Step 3: Compare the page length with the EL*

In preprocessing, the value of the EL is recorded by sending a set of wrong passwords. In this step, the total length of the returned page is compared to the EL. If they are equal, then we have reason to believe that the pair of passwords are wrong passwords, otherwise it can proceed to the next judgment. This step can filter out interference Events 1, 2, and 4.

*4) Step 4: Recheck*

With the above three steps, a set of passwords, i.e. $s$, is gotten to be detected. To ensure the accuracy of the results, the system will also perform a re-sending detection, which is called Recheck in this paper (refer to the recheck process in the supplemental material 2).

In this step, after a pre-request, the system will send an incorrect password $e_1$ and a set of passwords $s$ to be detected from the previous step. The returned page lengths of the two are compared. If they are equal, the web page may have been blocked by the system due to too many attempts being made. At this time, neither the correct password nor the wrong password is accepted by the system, and so the length of the returned value will be the same. This step is used to filter out interference Events 3 and 5.

*E. Response event discrimination*

When the input username or password is incorrect, the blasting failure response can be obtained through the above mentioned blasting result judgement. In order to analyze the causes of blasting failure and provide reasonable guidance for the next round of blasting, this paper classifies the causes of blasting failure into the following categories based on the keyword information of the response pages:

(1) Username does not exist or is incorrect. In this case, the system may prompt "the username does not exist" or even prompt "the username or password error" so as to obfuscate the failure causes.

(2) Password error (Event 2、Event4). After we enter the username and password on the login page, the login process compares the user password in the user account database. If the user password fails to match, the response page may display an error password message. This corresponds to Events 2 and 4.

Exceed the maximum number of password errors (Event3、Event5). In the blasting process, weak passwords are generally continuously entered for login test, and each wrong password will cause a login failure. To avoid blasting attacks, the system may set the maximum number of password errors. If the number of password errors exceeds this limitation, the system may prompt "Exceed the maximum number of password errors". This corresponds to Events 3 and 5.

(3) Firewall blocking (Event6). The firewall can filter incoming and outgoing data, manage incoming and outgoing behaviors, block some prohibited services, record information content and activities that pass through the firewall, and detect and alarm network attacks. When the firewall is enabled in the system, the user blasting test may trigger firewall blocking. At this time, the response page may prompt "web page is blocked" or other messages. This corresponds to Event 6.

(4) Other (Event1、Event7). Other cases, such as Event 1 "No login failure message" or Event 7 "Other".

In this section, keywords of the response page are extracted and matching rules are designed to divide the reasons for blasting failures. Combined with the reason of blasting failure, we can adjust the following weak password selection and blasting method. In addition, page features can also be extracted and reasons for blasting failure can be identified by machine learning method. We will leave this for further study.

F. *Username blasting based on response time*

When blasting a login box, guess the username at first and then search for the password for the existing username can greatly improve the blasting efficiency compared with blasting the username and password at the same time.

The existence of the username is usually determined by the output (the length of the response package) on the page after the login failure, for example, if the username exists it may prompt "Password error!", if the username does not exist it may prompt "Username is incorrect!".

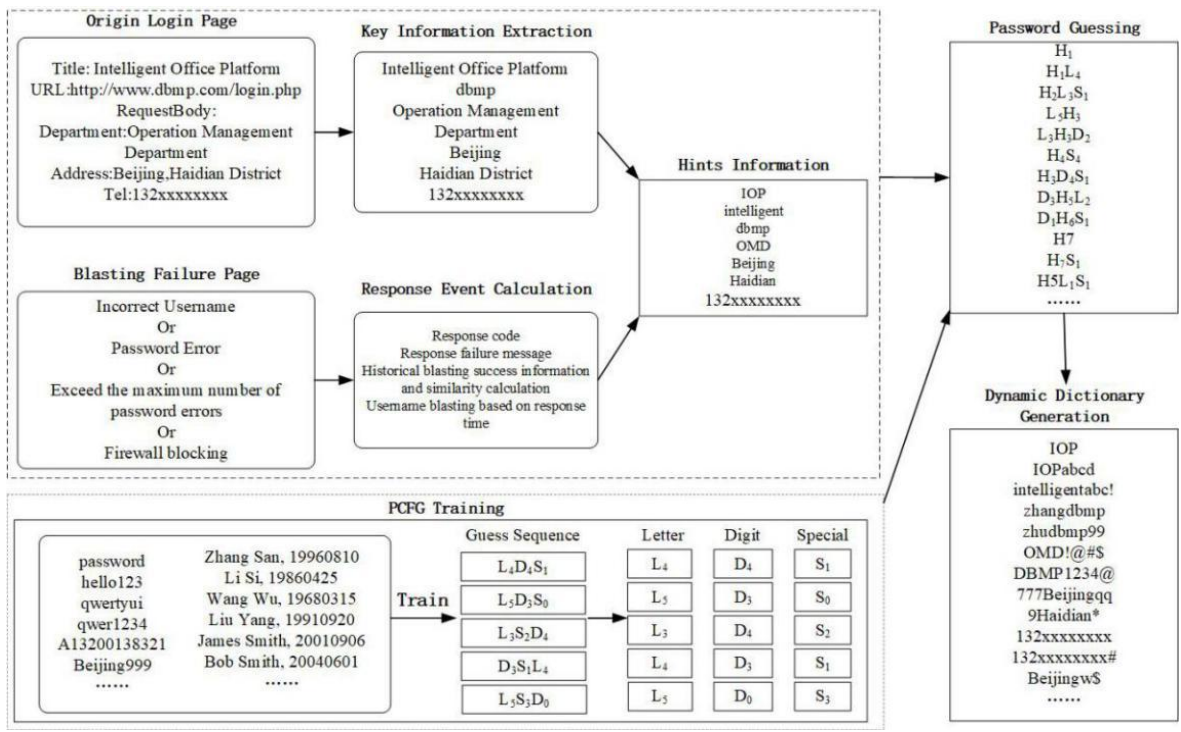

Fig. 2. Dynamic password dictionary adjustment process

However, if the system prompt "Username or password is incorrect!" in the login failure response page regardless of whether the username exists, is it impossible to determine the existence of the username in this case?

To try to guess a username in the case of obfuscation, we can determine the presence of a username based on the server response time. In detail, we can blast username with a very long password, sorted by response time, with the longest response time being the possible usernames.

According to the security specification, the database does not hold the plain text password, but the password hash value, which usually use the MD5 algorithm. Hashing is irreversible, it can only be encrypted but not decrypted, so we need to determine whether the password submitted by the user and the database is the same. We can encrypt the hashes submitted by the user and compare the hashes to see if they are the same. The hashing process takes time and longer passwords take more time for encryption.

When the username submitted by the user does not exist, the password cannot be found in the database and the program cannot go through the hashing step. However, when the user name is present, the program needs to hash, and we deliberately submit a long password to delay the time, resulting in a difference in server response time, based on this difference we can determine the existence of the username. This method may cause some errors, we can change the possible usernames in the dictionary for repeated tests, if the username is still prominent in the first place after several tests, then we can confirm the existence of the username.

*G. Dynamic dictionary adjustment*

After in-depth analysis and research, we found that in addition to extracting keyword information from the original login page, as introduced in Section IV, to generate a weak password dictionary, the results of each blasting event could also be used to guide and adjust the password type selection for the next round of blasting.

As shown in Fig. 2, the system also extracts key information and analyze features from the blasting failure page. We can use the response code and response information when failed to blast a login page, as well as blasting history, blasting success events with similarity calculation, blasting success information such as usernames, combined with the original login page to extract key information, these clues together constitute hints information.

At the same time, we explicit PCFG algorithm to train weak password generation rules on the public weak password set, and on the basis of prompt (H), we combine character table, number table and special character table to guess and generate weak password dictionary. L indicates alphabet letters, D indicates digital numbers, S indicates special characters, the corresponding subscript indicates the number of letters, digits or special characters, H indicates a hint messages, and the subscript of H indicates the index of hint keywords.

In this way, after each or multiple rounds of blasting failure, we can analyze and calculate the response event with the blasting response page, and extract key information to guide and adjust the dictionary that can be adopted for subsequent weak password blasting, so as to improve the efficiency of weak password blasting.

*H. Calculation of detection capability table and false positive rate*

Let $S$ be the set of events that occur in all probes, with $S = M + N$, where $M$ is the event that the correct password page is detected, and $P(M)$ is its probability, while $N$ is the event set of the remaining interference events, and $P(N)$ is its probability, where the sub-event is recorded as $N_i$, and $i$ is the event label.

Let $R$ be the event that the detection algorithm can exclude, and $Q(R)$ be its exclusion probability. Further, $R_i$ is the interference sub-event $i$ that can be eliminated,

so it is included in the formula of the false alarm rate:

$$P(mistake) = \sum_{i=1}^{i \leq max} P(N_i)(1 - Q(R_i))$$

Except for preprocessing, the interference event probabilities identified by steps 1–4 of the algorithm are $Q(A)$, $Q(B)$, $Q(C)$, and $Q(D)$, respectively. Where the sub-event probability is $Q(K_i), K \in \{A, B, C, D\}$. Here, $i$ is the event number, $i \in \{1,2,3,4,5,6,7\}$.

The detection capability of each step in the algorithm for these events is shown in Table I.

TABLE I. RANGE OF DETECTION CAPABILITIES FOR EACH STEP

| Event / Steps | Event 1 | Event 2 | Event 3 | Event 4 | Event 5 | Event 6 | Event 7 |
|---|---|---|---|---|---|---|---|
| A |   | √ | √ | √ | √ | √ |   |
| B | √ | √ | √ |   |   |   |   |
| C | √ | √ |   | √ |   |   |   |
| D |   |   | √ |   | √ |   |   |

There are overlaps in the interception ability of each step in the algorithm, which can improve the final accuracy. It can be concluded that the probability of each interference event being eliminated is:

$Q(R1) = Q(B1 \cup C1)$

$Q(R2) = Q(A2 \cup B2 \cup C2)$

$Q(R3) = Q(A3 \cup B3 \cup D3)$

$Q(R4) = Q(A4 \cup C4)$

$Q(R5) = Q(A5 \cup D5)$

$Q(R6) = Q(A6)$

In this algorithm, ideally there is the case:

$Q(B1), Q(C1), Q(B2), Q(B3), Q(C2), Q(C4), Q(D3), Q(D5) = 1$

Therefore:

$Q(R1), Q(R2), Q(R3), Q(R4), Q(R5) = 1$

So the final false positive formula is:

$P(mistake) = P(N6)(1 - Q(R6)) + P(N7)$

Because $P(N7)$ is an uncontrollable situation, in the DBKER algorithm, the determining factor affecting the false positive rate is $R6$. In other words, the blacklist keywords for the universal password firewall blocking page will directly determine the final false positive rate.

## IV. WEBCRACK DETECTION SYSTEM

Based on the above DBKER algorithm, a basic system called WebCrack is implemented, which consists of page analysis module, random headers module and dictionary generation module (refer to the overall framework of WebCrack in the supplemental material 3 and the general workflow in the supplemental material 4).

### A. Page analysis module

WebCrack adopts static page analysis method. After identifying the submission path, the data is directly submitted to the target address without rendering the page. Some websites verify that the user's input contains illegal characters on the front-end to prevent universal password or background injection attacks [24], so WebCrack does not render pages to execute Javascript scripts, thus can circumvent such defenses.

*1) Login page identification*

This step has two main effects:

(1) Verify whether the page is a real background login page. The system uses keywords and other characteristics to determine whether this page is real. If it is not the login page, the subsequent detection will not be performed.

(2) Exclude pages with verification codes. Since this system is only for verifying the practicability of the DBKER algorithm, and does not implement identification of the verification code, the system will automatically exit when a page containing a verification code is encountered.

*2) Submission path identification*

To achieve automatic detection, artificial simulation is needed to identify the login box and the submitted parameters [25]. In general, the most common authentication methods for web systems are based on Form or Ajax [9]. While most CMSs (Content Management Systems) use the form method, this system only realizes the path and parameter identification in the format of Form.

Firstly, the BeautifulSoup module is used to analyze and extract the form, and then the value of the action field is extracted for analysis to obtain the correct submission path address.

*3) Submission parameter identification*

After extracting the form, we need to find the corresponding username, password input box, and login button. For the username and password input box, this process is mainly based on keyword comparison. If the name value in the input tag contains keywords such as 'user', 'name', 'zhanghao', 'yonghu', 'email', or 'account', then this is marked as the username input box. Similarly, if the tag contains keywords such as 'pass', 'pw', or 'mima', it is marked as a password field.

To prevent cross-site request forgery (CSRF) [26] attacks, some websites often add hidden token fields for verification. Therefore, the system will traverse all key-value pairs containing a value in the extracted form to bypass the detection of the CSRF protection system.

In addition, some websites will set a reset button. When this key contains a value, the form will be initialized, and the account password sent will be ignored. So when the system traverses to the reset field, it will be removed from the dictionary.

*B. Random headers module*

To prevent hackers from guessing the backend password, many WAF (Web Application Firewall) and web management systems have a set limit on the number of submissions for the same IP within a period of time. When attempts exceed a certain number of times, the system will be locked and the user cannot continue to log in. However, some developers use the X-Forwarded-For field in the request header to obtain the target IP, and as this field can be forged, a random function can be added to bypass this limitation. In WebCrack, a random set of User-Agent, X-Forwarded-For, and Client-IP fields are generated before each packet being sent to bypass the protection restrictions of WAF and CMS.

*C. Dictionary generation module*

The dictionary generation module is composed of three sub-dictionaries: general dictionary, dynamic dictionary, and universal password dictionary.

*1) General dictionary*

In addition to the regular dictionaries of 123456 and qwe123, WebCrack can also generate corresponding dictionaries based on the user name currently being probed. For example, when the detected user name is admin, the password associated with it is automatically generated, such as admin123, admin888, admin123456, and so on.

*2) Dynamic dictionary*

WebCrack can also generate different dynamic dictionaries based on PCFG algorithm according to the key information extracted from the origin login page, such as domain names or organization. For convenience, many administrators will set their own management password according to the domain name, which can be easily guessed by hackers. If the system detects a domain name, for example webcrack.yzddmr6.com, the dynamic dictionary list will be generated, where the suffix can be configured by the user; although, if the regular match is an IP instead of a domain name, no dynamic dictionary will be generated (refer to the supplemental material 5).

In addition, WebCrack can also adjust the dynamic dictionaries via response event discrimination and analysis, to provide hint information and constructive suggestions for more effective blasting.

*3) Universal password dictionary*

A major feature of WebCrack is that it can support the detection of universal passwords. There are some common payloads for detecting universal passwords, such as admin 'or' a '=' a, 'or' = 'or', admin 'or' 1 '=' 1 'or 1 = 1,') or ('a' = 'a, etc. The system thus first performs a traditional dictionary attempt, and if the correct password is not found, the universal password detection module will be enabled.

Since most mature management systems do not have this vulnerability. Then enabling this option for all targets will cause unnecessary packet sending and can even trigger the firewall blocking. So the advantage of custom blasting rules becomes apparent. For an unknown CMS, the universal password detection is enabled by default; for a known one, whether to enable this can be set via custom rules (refer to configuration description in the supplemental material 6).

## V. EXPERIMENTS AND EVALUATION ANALYSIS

### A. Experiment 1: DedeCMS Test

In this experiment, the widely used dream weaving management system DedeCMS[3] is chosen. To verify the reliability of the algorithm, the proposed DBKER algorithm is used in experiments 1 and 2 for testing, and the custom blasting mode is disabled.

*1) Experimental environment*

The experimental environment is shown in Table II.

| Program Version | DedeCMS V57_GBK_SP2 |
|---|---|
| Server system and PHP version | WINNT / PHP v5.6.27 |
| Server software | PHP 5.6.27 Development Server |
| Server MySQL version | 5.5.53 |

TABLE II. DEDECMS EXPERIMENTAL ENVIRONMENT

*2) Test and analysis*

At first, set the password as admin: yzddmr6123, and then enter the background address into WebCrack for detection.

We use BurpSuite [27] software here to analyze the data packets sent by the system and the change in the length of the return value of the data packet, as shown in Fig. 3.

The No.1 data packet first sent a pre-request. At this time, the CMS assigns a cookie to the user. The response page contains the set-cookie field, so the length of the first data packet is different.

The data packets with the serial numbers 2 and 3 sent wrong passwords, and the length of return values of the two were the same. The system considers that this page is stable and has the conditions for blasting. The EL at this moment was recorded as 1490. After receiving the wrong password, this management system will prompt the wrong password on the new page, which belongs to event 4. The first and third steps of the algorithm can filter this event (refer to the return information of packet 2 during DedeCMS Test in the supplemental material 7).

---

[3] DedeCMS Official Website: http://www.dedecms.com

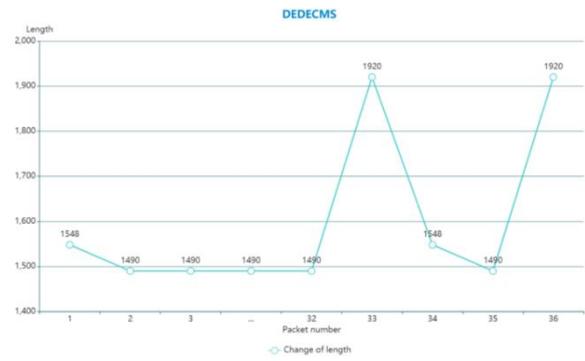

Fig. 3. Changes in the length of the DedeCMS packet return value

The data packet with the serial number 33 sent the correct password, which caused the length of the page return value to change. After receiving the correct password, the management system prompted that the password is correct on a new page, and event 8 occurs. At this point, the keyword blacklist detection was passed. At the same time, the key of the login box did not exist, and the length of the return value was not equal to the EL, so it entered the Recheck (refer to the return information in the supplemental material 8).

In the packet No.34, it shows that a pre-request was first performed in the Recheck link. Then the 35th data packet sent another wrong password, and the length of this return value was recorded as 1490.

The 36th data packet sent the correct password. At this time, event 8 occurred, and the length of the returned value changed to 1920, which was not equal to the 1490 of the 35th data packet. At this time, the final correct password was yzddmr6123 (refer to the progress in the supplemental material 9).

*3) False positive test*

The program not only is able to correctly identify weak passwords, but also does not produce false positives for systems that do not have weak passwords.

At this time, the password was modified to a strong password that the program cannot detect, i.e., eaa4a6d7a3ed765985758796f13bd26a, and a false positive test of the program was performed (refer to the progress in the supplemental material 10).

B. *Experiment 2: Discuz! Test*

Discuz![4] is a universal community forum software system, and it is one of the most mature forum software systems in the world.

*1) Experimental environment*

The experimental environment is shown in Table III.

TABLE III. DISCUZ! EXPERIMENTAL ENVIRONMENT

| Program Version | Discuz! X3.2 Release 20140618 |
|---|---|
| Server system and PHP version | WINNT / PHP v5.6.27 |
| Server software | PHP 5.6.27 Development Server |
| Server MySQL version | 5.5.53 |

*2) Testing and analysis*

The password was set in the management console as admin: admin888, and then the BurpSuite software was used to capture the package drawing, as shown in Fig. 4.

In packet number 37, WebCrack first sent a pre-request. Packets 38 and 39 sent the wrong password and the EL was recorded. From the return information, it was found that the system continues to display the login box on the original page in the face of the wrong password, and there was no prompt, and so event 1 occurs.

When sending the No.42 data packet, the correct password was sent. At this time, the system jumped to 302 and jumped to the background page. Since we only care about the page after the 302 jump, at this time we checked the content of packet 43. Event 9 occurred at this time and entered the management background.

---

[4] Discuz! Official Website - PHP open source forum: https://www.discuz.net/forum.php

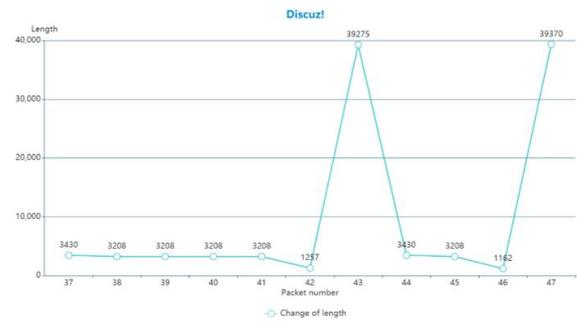

Fig. 4. Change of length of the Discuz! packet return value

Then the system entered the Recheck. At this point, the return value of the 45th packet was different from the 47th packet. Through Recheck detection, the final correct password admin888 was obtained (refer to the supplemental material 11).

*3) False positive test*

At this time, the password was changed to a strong password that the program cannot detect, i.e., eaa4a6d7a3ed765985758796f13bd26a, to perform a false positive test of the program. WebCrack had no false positives in the face of strong passwords (refer to the supplemental material 12).

C. *Experiment 3: Random back-end comparison test*

*1) Data selection*

In order to verify the actual effect of weak password detection, we randomly crawled 14185 possible web background addresses and detected them using WebCrack and web_pwd_common_crack, respectively.

*2) Experimental results*

As a weak password detection tool, only accuracy and recognition rate are mainly considered (refer to detailed description in the supplemental material 13). The statistical results after calculating the accuracy and recognition rate are shown in Table IV.

*3) Data analysis*

*a) Comparative analysis*

In terms of accuracy, WebCrack is 38.39 percentage points higher than web_pwd_common_crack. From the perspective of the recognition rate, WebCrack's recognition rate is about 4 times than that of web_pwd_common_crack.

TABLE IV. STATISTICS OF THE EXPERIMENTAL RESULTS

| Metrics | WebCrack | web_pwd_common_crack |
|---|---|---|
| n(success) | 80 | 56 |
| n(fail) | 1014 | 1804 |
| n(effect) | 75 | 31 |
| n(wrong) | 5 | 25 |
| n(error) | 6706 | 3607 |
| p(correct) | 93.75% | 55.36% |
| p(recognize) | 6.86% | 1.67% |

Judging from the timeout and the number of blocked logs, WebCrack is almost twice that of web_pwd_common_crack. The reason for this is that although the universal password recognition module has increased the recognition rate, the number of times it is intercepted by various firewalls has increased as it sends a large number of payloads containing attacks in a short period of time. In the later stage, many websites may be inaccessible because the IP is blocked by the firewall, so fewer pages are identified. When the number of identified pages is less than with web_pwd_common_crack, the number, the accuracy, and the recognition rate of correct weak passwords detected by WebCrack are still higher than with web_pwd_common_crack, which shows that WebCrack has better practicability.

*b) Weak password composition*

After analysis, among the 75 valid weak passwords detected by WebCrack, 38 of them are universal passwords, 23 of them are admin/admin, 5 of them are admin/123456, 3 of them are admin/admin888, and 6 of them are other weak passwords. The overall composition is shown in Fig. 5.

It is shown that the proportion of universal password vulnerabilities has exceeded 50%, this inspire us to pay more attention to the protection of the universal password vulnerabilities.

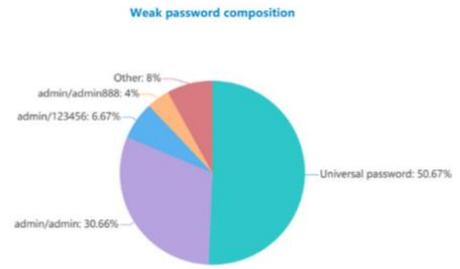

Fig. 5. Composition of the weak passwords in the detection results

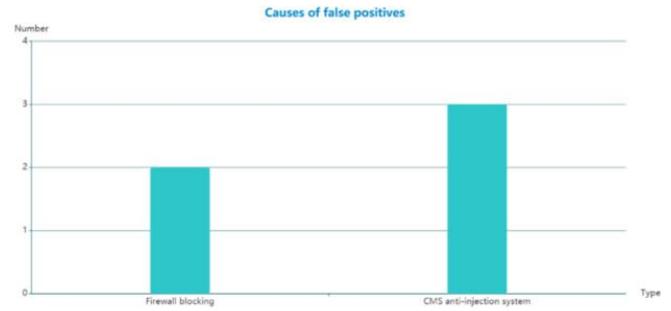

Fig. 6. Reasons for the false positives

*c) Reasons for the false positives*

The reasons for false positives are summarized and sorted out, as shown in Fig. 6. After debugging and analysis, it is found that the main causes of false positives are all from the universal password detection module.

When the firewall or CMS interception system is triggered, the length of the return value changes, and event 6 will occur. Here, the system cannot tell whether the background page is entered or blocked by the firewall. If the features of the interception system are not in the blacklist, it will be misjudged as a page that successfully entering the background page. Hence, the result accords with the conclusion of the false positive rate formula.

VI. SUMMARY

Vulnerabilities such as weak passwords and universal passwords are relatively simple and easy to be exploited, which require no high-level technical skills. If used by criminals, they may lead to leakage of personal information and even losses of server authorities, which could cause huge loss to the government, enterprises and individuals. In this paper, we proposed an effective algorithm named DBKER for web weak password and universal password automatic detection, then a specific system WebCrack is implemented on the basis of this algorithm. Batch detection can be carried out by simply importing the backend address, and it exceeds the benchmark systems on the market in terms of accuracy and detection rate.

SUPPLEMENTARY MATERIAL

Supplementary material files are described in this section. File 1 and file 2 show the pre-processing and the recheck procedure of DBKER blasting algorithm respectively. File 3 shows the overall framework of the WebCrack system, which includes the page analysis module, the random headers module and the dictionary generation module. File 4 shows the general workflow of the WebCrack system. File 5 takes "webcrack.yzddmr6.com" as an example to show the dynamic dictionary generation procedure. File 6 describes the rules for custom judgement. File 7 shows the return information of packet 3 during DedeCMS Test. File 8 shows the change of page length after sending the correct password. File 9 shows the response of WebCrack after detecting the correct password during the DedeCMS Test and Discuz! Test. File 10 and file 12 show the response of false positive test. File 11 shows the response of recheck detection. File 13 describes the evaluation indicators including accuracy p(correct) and recognition rate p(recognize).

ACKNOWLEDGMENTS

This paper is supported by National Natural Science Foundation of China: 62127808, 62172176, 62072200.

# Supplemental Files

1. The pre-processing procedure is shown below. In detail, the system will first send two incorrect passwords to determine the stability of the page. If the lengths of the two returned pages are equal, the page is considered stable. Meanwhile, the total length of the page returned at this time, i.e. the Error Length (hereinafter referred to as EL), is recorded as the benchmark for the judgment in the third step. If unstable, the program exits.

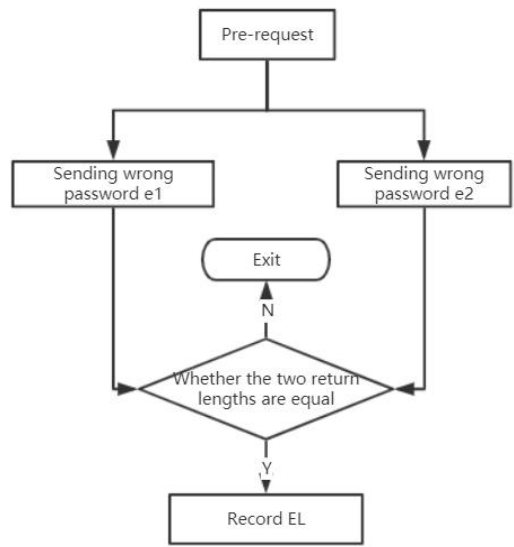

2. The recheck process is shown below. In this step, after a pre-request, the system will send an incorrect password $e_1$ and a set of passwords s to be detected from the previous step. The returned page lengths of the two are compared. If they are equal, the web page may have been blocked by the system due to too many attempts being made. At this time, neither the correct password nor the wrong password is accepted by the system, and so the length of the returned value will be the same. This step is used to filter out interference events 3 and 5.

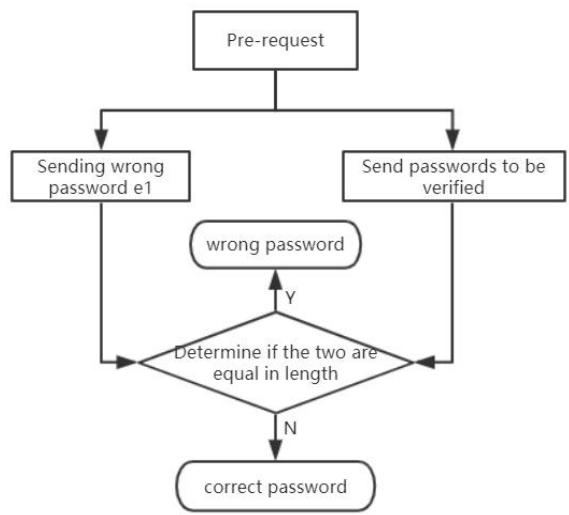

3. The overall framework of the WebCrack system is shown below, which includes the page analysis module, the random headers module and the dictionary generation module.

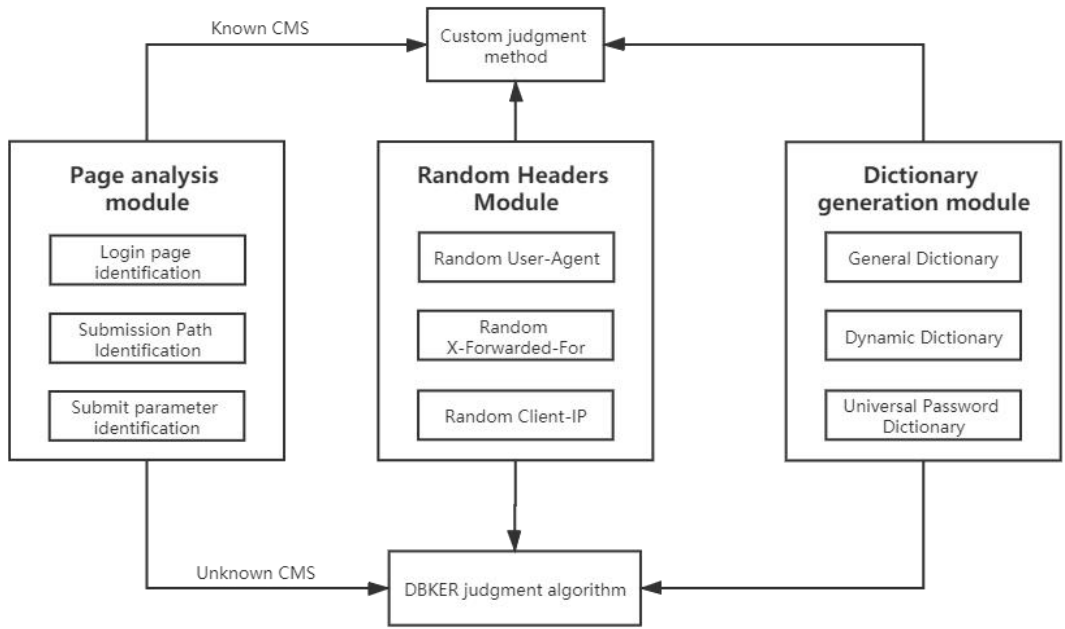

4. The general workflow of WebCrack system is shown below:

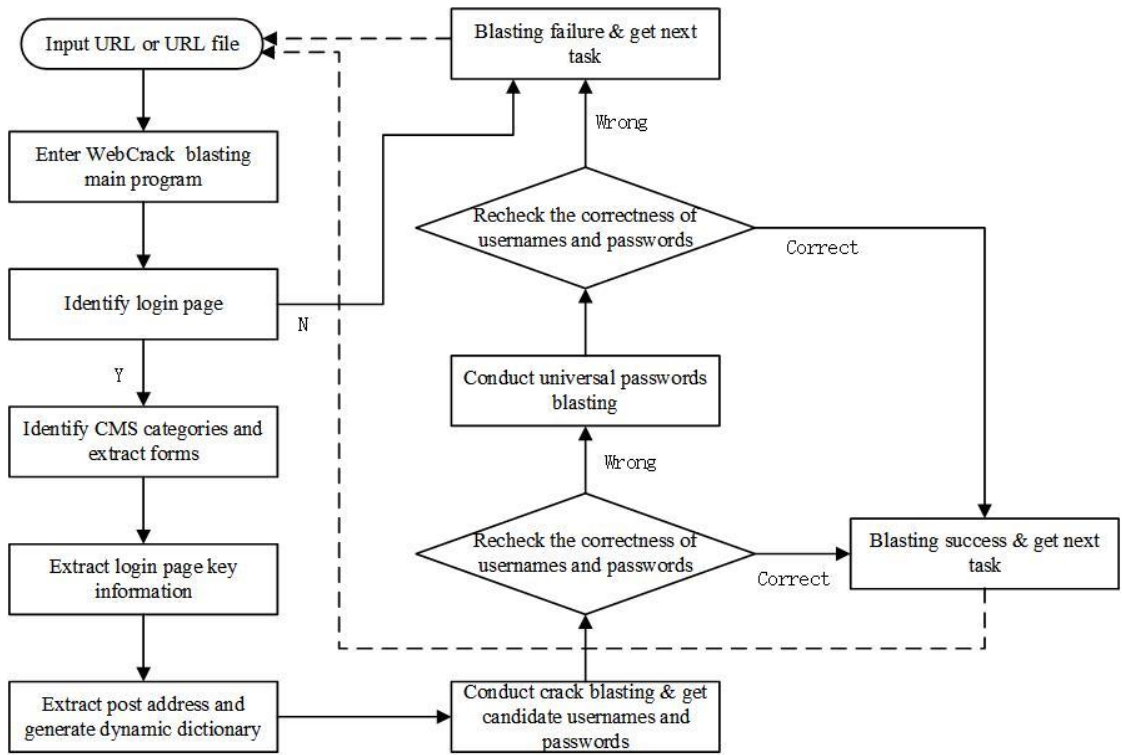

5. If the system detects a domain name, for example webcrack.yzddmr6.com, the following dynamic dictionary list will be generated, where the suffix can be configured by the user; although, if the regular match is an IP instead of a domain name, no dynamic dictionary will be generated.

```
webcrack.yzddmr6.com
```

```
yzddmr6.com
webcrack
webcrack123
webcrack888
webcrack666
webcrack123456
yzddmr6
yzddmr6123
yzddmr6888
yzddmr6666
yzddmr6123456
```

6. With the universal weak password determination algorithm, although the system can blast most website backends, there will always be some special websites that are not standardized or that do not meet the general rules. To solve this case, the system has added a custom judgment rule module, and users can customize the rules according to their needs.

Configuration file parameter description:

```
[
   {
      "name":"Name of the cms",
      "keywords":" Keywords of the cms ",
      "captcha":"1 is the verification code in the background, 0 is not. Because this version does not process the verification code, so 1 will exit the blasting ",
      "exp_able":" Whether to enable the universal password module blasting ",
      "success_flag":" Keywords of the page after successful login ",
      "fail_flag":" If you fill in this item, it will exit the blasting when it encounters the keywords in it. It is used to limit the number of blasting cms. ",
      "alert":" If it is 1, the contents of the note below will be printed.",
      "note":" Please ensure that this file is in UTF-8 format."
   }
]
```

7. Return information of packet 2 during DedeCMS Test.

8. The page length changed after sending the correct password:

9. WebCrack detected the correct password during DedeCMS Test and Discuz! Test.

```
$ python3 webcrack.py

*****************************************************
*                                                   *
***************     Code By yzddMr6     *************
*                                                   *
*****************************************************
File or Url:
http://yzddmr6.com:9090/cms/DedeCMS-V5.7-GBK-SP2/uploads/dede/login.php
Checking : http://yzddmr6.com:9090/cms/DedeCMS-V5.7-GBK-SP2/uploads/dede/login.php 2020-02-23 19:56:42
字典总数:  33  当前尝试:   1  checking: admin admin
字典总数:  33  当前尝试:   2  checking: admin 123456
字典总数:  33  当前尝试:   3  checking: admin admin888
字典总数:  33  当前尝试:   4  checking: admin 12345678
字典总数:  33  当前尝试:   5  checking: admin 123123
字典总数:  33  当前尝试:   6  checking: admin 88888888
字典总数:  33  当前尝试:   7  checking: admin 888888
字典总数:  33  当前尝试:   8  checking: admin password
字典总数:  33  当前尝试:   9  checking: admin 123456a
字典总数:  33  当前尝试:  10  checking: admin admin123
字典总数:  33  当前尝试:  11  checking: admin admin123456
字典总数:  33  当前尝试:  12  checking: admin admin666
字典总数:  33  当前尝试:  13  checking: admin admin2018
字典总数:  33  当前尝试:  14  checking: admin 123456789
字典总数:  33  当前尝试:  15  checking: admin 654321
字典总数:  33  当前尝试:  16  checking: admin 666666
字典总数:  33  当前尝试:  17  checking: admin 66666666
字典总数:  33  当前尝试:  18  checking: admin 1234567890
字典总数:  33  当前尝试:  19  checking: admin 8888888
字典总数:  33  当前尝试:  20  checking: admin 987654321
字典总数:  33  当前尝试:  21  checking: admin 0123456789
字典总数:  33  当前尝试:  22  checking: admin 12345
字典总数:  33  当前尝试:  23  checking: admin 1234567
字典总数:  33  当前尝试:  24  checking: admin 000000
字典总数:  33  当前尝试:  25  checking: admin 111111
字典总数:  33  当前尝试:  26  checking: admin 5201314
字典总数:  33  当前尝试:  27  checking: admin 123123
字典总数:  33  当前尝试:  28  checking: admin yzddmr6.com
字典总数:  33  当前尝试:  29  checking: admin yzddmr6
字典总数:  33  当前尝试:  30  checking: admin yzddmr6123
Rechecking... http://yzddmr6.com:9090/cms/DedeCMS-V5.7-GBK-SP2/uploads/dede/login.php admin yzddmr6123
[+] Success : http://yzddmr6.com:9090/cms/DedeCMS-V5.7-GBK-SP2/uploads/dede/login.php  user/pass admin/yzddmr6123
```

10. As can be seen from the figure below, WebCrack had no false positives.

```
字典总数:  33  当前尝试:  29  checking: admin yzddmr6
字典总数:  33  当前尝试:  30  checking: admin yzddmr6123
字典总数:  33  当前尝试:  31  checking: admin yzddmr6888
字典总数:  33  当前尝试:  32  checking: admin yzddmr6666
字典总数:  33  当前尝试:  33  checking: admin yzddmr6123456
Exp_dic is trying
字典总数:  25  当前尝试:   1  checking: admin' or 'a'='a admin' or 'a'='a
字典总数:  25  当前尝试:   2  checking: admin' or 'a'='a 'or'='or'
字典总数:  25  当前尝试:   3  checking: admin' or 'a'='a admin' or '1'='1' or 1=1
字典总数:  25  当前尝试:   4  checking: admin' or 'a'='a ')or('a'='a
字典总数:  25  当前尝试:   5  checking: admin' or 'a'='a 'or 1=1 -- -
字典总数:  25  当前尝试:   6  checking: 'or'='or' admin' or 'a'='a
字典总数:  25  当前尝试:   7  checking: 'or'='or' 'or'='or'
字典总数:  25  当前尝试:   8  checking: 'or'='or' admin' or '1'='1' or 1=1
字典总数:  25  当前尝试:   9  checking: 'or'='or' ')or('a'='a
字典总数:  25  当前尝试:  10  checking: 'or'='or' 'or 1=1 -- -
字典总数:  25  当前尝试:  11  checking: admin' or '1'='1' or 1=1 admin' or 'a'='a
字典总数:  25  当前尝试:  12  checking: admin' or '1'='1' or 1=1 'or'='or'
字典总数:  25  当前尝试:  13  checking: admin' or '1'='1' or 1=1 admin' or '1'='1' or 1=1
字典总数:  25  当前尝试:  14  checking: admin' or '1'='1' or 1=1 ')or('a'='a
字典总数:  25  当前尝试:  15  checking: admin' or '1'='1' or 1=1 'or 1=1 -- -
字典总数:  25  当前尝试:  16  checking: ')or('a'='a admin' or 'a'='a
字典总数:  25  当前尝试:  17  checking: ')or('a'='a 'or'='or'
字典总数:  25  当前尝试:  18  checking: ')or('a'='a admin' or '1'='1' or 1=1
字典总数:  25  当前尝试:  19  checking: ')or('a'='a ')or('a'='a
字典总数:  25  当前尝试:  20  checking: ')or('a'='a 'or 1=1 -- -
字典总数:  25  当前尝试:  21  checking: 'or 1=1 -- - admin' or 'a'='a
字典总数:  25  当前尝试:  22  checking: 'or 1=1 -- - 'or'='or'
字典总数:  25  当前尝试:  23  checking: 'or 1=1 -- - admin' or '1'='1' or 1=1
字典总数:  25  当前尝试:  24  checking: 'or 1=1 -- - ')or('a'='a
字典总数:  25  当前尝试:  25  checking: 'or 1=1 -- - 'or 1=1 -- -
[-] Faild : http://yzddmr6.com:9090/cms/DedeCMS-V5.7-GBK-SP2/uploads/dede/login.php 2020-02-23 20:49:50
```

11. Through Recheck detection, the final correct password admin888 was obtained.

```
$ python3 webcrack.py
***************************************************
*                                                 *
***************    Code By yzddMr6    *************
*                                                 *
***************************************************
File or Url:
http://yzddmr6.com:9090/cms/discuz_x3.2_sc_utf8_0618/upload/admin.php
Checking : http://yzddmr6.com:9090/cms/discuz_x3.2_sc_utf8_0618/upload/admin.php 2020-02-23 20:08:43
字典总数:  33  当前尝试:   1  checking: admin admin
字典总数:  33  当前尝试:   2  checking: admin 123456
字典总数:  33  当前尝试:   3  checking: admin admin888
Rechecking... http://yzddmr6.com:9090/cms/discuz_x3.2_sc_utf8_0618/upload/admin.php admin admin888
[+] Success : http://yzddmr6.com:9090/cms/discuz_x3.2_sc_utf8_0618/upload/admin.php  user/pass admin/admin888
```

12. WebCrack had no false positives in the face of strong passwords:

```
字典总数:  33  当前尝试:  31  checking: admin yzddmr6888
字典总数:  33  当前尝试:  32  checking: admin yzddmr6666
字典总数:  33  当前尝试:  33  checking: admin yzddmr6123456
Exp_dic is trying
字典总数:  25  当前尝试:   1  checking: admin' or 'a'='a admin' or 'a'='a
字典总数:  25  当前尝试:   2  checking: admin' or 'a'='a 'or'='or'
字典总数:  25  当前尝试:   3  checking: admin' or 'a'='a admin' or '1'='1' or 1=1
字典总数:  25  当前尝试:   4  checking: admin' or 'a'='a ')or('a'='a
字典总数:  25  当前尝试:   5  checking: admin' or 'a'='a 'or 1=1 -- -
字典总数:  25  当前尝试:   6  checking: 'or'='or' admin' or 'a'='a
字典总数:  25  当前尝试:   7  checking: 'or'='or' 'or'='or'
字典总数:  25  当前尝试:   8  checking: 'or'='or' admin' or '1'='1' or 1=1
字典总数:  25  当前尝试:   9  checking: 'or'='or' ')or('a'='a
字典总数:  25  当前尝试:  10  checking: 'or'='or' 'or 1=1 -- -
字典总数:  25  当前尝试:  11  checking: admin' or '1'='1' or 1=1 admin' or 'a'='a
字典总数:  25  当前尝试:  12  checking: admin' or '1'='1' or 1=1 'or'='or'
字典总数:  25  当前尝试:  13  checking: admin' or '1'='1' or 1=1 admin' or '1'='1' or 1=1
字典总数:  25  当前尝试:  14  checking: admin' or '1'='1' or 1=1 ')or('a'='a
字典总数:  25  当前尝试:  15  checking: admin' or '1'='1' or 1=1 'or 1=1 -- -
字典总数:  25  当前尝试:  16  checking: ')or('a'='a admin' or 'a'='a
字典总数:  25  当前尝试:  17  checking: ')or('a'='a 'or'='or'
字典总数:  25  当前尝试:  18  checking: ')or('a'='a admin' or '1'='1' or 1=1
字典总数:  25  当前尝试:  19  checking: ')or('a'='a ')or('a'='a
字典总数:  25  当前尝试:  20  checking: ')or('a'='a 'or 1=1 -- -
字典总数:  25  当前尝试:  21  checking: 'or 1=1 -- - admin' or 'a'='a
字典总数:  25  当前尝试:  22  checking: 'or 1=1 -- - 'or'='or'
字典总数:  25  当前尝试:  23  checking: 'or 1=1 -- - admin' or '1'='1' or 1=1
字典总数:  25  当前尝试:  24  checking: 'or 1=1 -- - ')or('a'='a
字典总数:  25  当前尝试:  25  checking: 'or 1=1 -- - 'or 1=1 -- -
[-] Faild : http://yzddmr6.com:9090/cms/discuz_x3.2_sc_utf8_0618/upload/admin.php 2020-02-23 20:59:53
```

13. Evaluation indicators

As a weak password detection tool, only two aspects of data are mainly considered:

(1) Accuracy: p(correct)

That is, among the results n (success) successfully identified by the tool, the number of effective weak passwords n (effect) compared to the number of successful identifications n(success). This result mainly reflects the accuracy of the tool's weak password vulnerability detection results, calculated as follows：

$$p(correct) = \frac{n(effect)}{n(success)}$$

(2) Recognition rate: p(recognize)

That is, the ratio of the number of identified effective weak passwords n(effect) to all detectable pages n(success + failure). This result mainly reflects the degree to which the tool recognizes weak

password vulnerabilities, calculated as follows:

$$p(\text{recognize}) = \frac{n(\text{effect})}{n(\text{success} + \text{fail})}$$